\documentclass[9pt,twocolumn,twoside]{osajnl}

\usepackage{amsmath}
\usepackage{amssymb}

\journal{ol} 

\setboolean{shortarticle}{true}

\title{On z-coherence of beams radiated by Schell-model sources with Gaussian profile}

\author[1,*]{M. Santarsiero}
\author[2]{G. Piquero}
\author[3]{J. C. G. de Sande}
\author[4]{O. Korotkova}
\author[2]{R. Mart\'inez-Herrero}
\author[1]{F. Gori}

\affil[1]{Dipartimento di Ingegneria Industriale, Elettronica e Meccanica, Universit\`a Roma Tre, Via V. Volterra 62, 00146 Rome, Italy}
\affil[2]{Departamento de \'{O}ptica, Universidad Complutense de Madrid, Ciudad Universitaria,  28040 Madrid, Spain}
\affil[3]{ETSIS de Telecomunicaci\'{o}n, Universidad Polit\'{e}cnica de Madrid, Campus Sur 28031 Madrid, Spain}
\affil[4]{Department of Physics, University of Miami, Coral Gables, FL, USA}
\affil[*]{Corresponding author: msantarsiero@uniroma3.it
\newline
\newline
\bf © 2022 Optica Publishing Group. One print or electronic copy may be made for personal use only. Systematic reproduction and distribution, duplication of any material in this paper for a fee or for commercial purposes, or modifications of the content of this paper are prohibited.
\newline
\url{https://doi.org/10.1364/OL.458764}
}

\begin{abstract}
The degree of coherence and the intensity distribution on the axis of the beam radiated by a planar partially coherent source of the Schell-model type are investigated. We present an expression for the on-axis cross-spectral density which is valid for a very general Schell-model source, with the only constraint that the intensity distribution across the source is Gaussian. Furthermore, we show that such an expression takes very simple analytical forms for several commonly used degrees of coherence of the source.
\end{abstract} 

\setboolean{displaycopyright}{true}

\begin{document}             

\maketitle     

The second-order coherence properties of a partially coherent source, as well as those of the beam radiated from it, are described either by its mutual coherence function or by its cross-spectral density (CSD) function~\cite{ManWolf95}, depending on whether the space-time or the space-frequency domain is considered, the two functions being related by a temporal Fourier transform. 

When dealing with the correlations between points lying along the propagation axis of a beam, two different aspects have to be taken into account. One is related to the finite bandwidth of the (temporal-)frequency spectrum of the radiating source. In particular, due to the Wiener-Kintchine theorem~\cite{ManWolf95}, a finite-bandwidth spectrum causes lack of correlation between points that are more than a coherence length apart. The latter is directly related to the coherence time, which, in turn, is given by the inverse of the spectrum bandwidth, and is a manifestation of the so-called temporal coherence.

The second aspect is purely spatial and only depends on the spatial-coherence properties of the source. It can be observed even when the distance between the points is well within the coherence length of the radiation, and is often referred to as \emph{longitudinal spatial coherence}~\cite{Rosen1995, Ryabukho2006, Cai2012}. Nonetheless, to avoid confusion between denominations we prefer to use the name \emph{$z$-coherence}~\cite{Gori2022} for this kind of correlation.
The most proper tool to investigate $z$-coherence is the CSD, because it refers to a specific temporal frequency and then to the monochromatic regime. 

Both these effects affect the longitudinal coherence properties of a beam~\cite{Ryabukho2009, Ryabukho2013} and have to be considered whenever such properties are used in applications based on it. This is the case, just to quote some, of optical coherence tomography~\cite{Abdulhalim2012, Ahmad2019}, profilometry~\cite{Rosen1996, Rosen2000}, interferometry~\cite{Ryabukho2004}, and ghost imaging~\cite{Cai2005, Liu2008, Magatti2009}.

Despite the broad interest for problems related to the longitudinal spatial coherence of a beam, the results presented so far mainly concern specific type of sources, such as spatially incoherent, homogeneous and quasi-homogeneous ones. On the contrary, it seems that the subject has not been dealt with yet for the wide class of the Schell-model sources, if not for the celebrated example of the Gaussian Schell-model (GSM) sources~\cite{ManWolf95, Friberg1983, Cai2005}, although they may represent a useful model for many physically realizable partially coherent sources.

In this letter we derive a simple expression for the CSD of the beam radiated by a general Schell-model (SM) source~\cite{ManWolf95} for points along the axis, with the only constraint that the intensity profile of the source is Gaussianly shaped and the paraxial conditions are met. The obtained expression provides the spatial correlation between two axial points, as well as the on-axis intensity, and in several cases of commonly used degrees of coherence reduces to a very simple analytical form.

\vspace{0.2cm}

We consider a Schell-model source across the plane $z=0$. For simplicity, we shall limit ourselves to the one-dimensional case. The obtained results, of course, can be extended at once to the two-dimensional case with rectangular symmetry. Under such hypotheses the CSD of the source takes the form
\begin{equation}
W_0(\xi_1,\xi_2)=\tau^*(\xi_1) \tau(\xi_2) \mu(\xi_1-\xi_2)
\; ,
\label{W0}
\end{equation}
where $\xi$ denotes the coordinate across the source, $\tau$ is a generic function giving account of the source profile and $\mu$ is the degree of coherence (DOC). We can express the latter through its Fourier transform~\cite{Bracewell2000}, namely,
\begin{equation}
\mu(\xi)
=
 \int_{-\infty}^\infty p(v) \; e^{- 2 \pi {\rm i} \xi v} {\rm d} v
\; ,
\label{W0_0.5}
\end{equation}
so that the CSD becomes
\begin{equation}
W_0(\xi_1,\xi_2)=\tau^*(\xi_1) \tau(\xi_2) \int_{-\infty}^\infty p(v) e^{- 2 \pi {\rm i} (\xi_1-\xi_2) v} {\rm d} v
\; .
\label{W0_1}
\end{equation}
We know that, due to the Bochner theorem, the function $p$ has to be nonnegative in order for the CSD to be \emph{bona fide}~\cite{ManWolf95}.

Equation~(\ref{W0_1}) can be viewed as a pseudo-modal expansion of the CSD~\cite{Gori2007,MartinezHerrero2009}. In fact, we have
\begin{equation}
W_0(\xi_1,\xi_2)
= 
\displaystyle \int_{-\infty}^\infty p(v) \,
\psi_0^*(v,\xi_1) \, 
\psi_0(v,\xi_2) \;
{\rm d} v
\; .
\label{W0_2}
\end{equation}
with the pseudomodes given by
\begin{equation}
\psi_0(v,\xi)
=  
\tau(\xi) \, e^{2 \pi {\rm i} \xi v}
\; .
\label{W0_2.5}
\end{equation}

To evaluate the CSD of the radiated beam we can use the expression of the propagated modes $\psi_z(v,x)$, so that 
\begin{equation}
W(x_1,z_1; x_2, z_2)= \displaystyle \int_{-\infty}^\infty p(v) \,
\psi_{z_1}^*(v,x_1) \, 
\psi_{z_2}(v,x_2) \;
{\rm d} v
\; .
\label{W0_3}
\end{equation}
where, according to the Fresnel integral~\cite{BornWolf}
\begin{equation}
\psi_z(v,x) = \sqrt{\displaystyle\frac{- {\rm i}}{\lambda z}} \; e^{{\rm i} k z}
\displaystyle \int_{-\infty}^\infty \psi_0(v,\xi)
\exp\left[ \frac{{\rm i} k}{2 z} (x-\xi)^2\right]  \,
{\rm d} \xi
\; .
\label{W0_4}
\end{equation}
It is worth recalling that such an integral defines the so called Fresnel transform, for which a sampling theorem exists~\cite{Gori1981}. The latter could be of potential interest for the evaluation of the propagation of fields with finite support.

We can give $\psi_z(v,x)$ a closed form if we take the amplitude $\tau $ as Gaussian, i.e.,
\begin{equation}
\tau(\xi) = A \; e^{- (\xi/w_0)^2}
\; ,
\label{W0_5}
\end{equation}
with $A$ an amplitude factor and $w_0$ a positive constant, and use the form of the modes in Eq.~(\ref{W0_2.5}). In such a case we have~\cite{Santarsiero1996}
\begin{equation}
\psi_z(v,x) =
\displaystyle\frac{A \, e^{{\rm i} k z}
\, e^{{{\rm i}kx^2}/{(2z)}}
}{\sqrt{1 + {\rm i} \, z/ L}}
\;
\exp\left[
- \; \displaystyle\frac
{ k L \left(x-2 \pi v z/k  \right)^2}
{2z(z - {\rm i} L)}
\right]
\; ,
\label{W0_6}
\end{equation}
with $L=k w_0^2/2$ being the Rayleigh distance pertinent to a coherent Gaussian beam having waist size $w_0$~\cite{Siegman86}.

Although the above equations allow us to evaluate the propagated CSD at any point in space, they simplify considerably if we limit ourselves to the points on the $z$-axis (i.e., $x=0$), in which case the expression of the propagated modes becomes
\begin{equation} 
\psi_z(v,0) =
\displaystyle\frac{A \, e^{{\rm i} k z}}
{\sqrt{1 + {\rm i} \, z/ L}}
\;
\exp\left[ 
- \; \displaystyle\frac{2 \pi^2  L \, z }{k(z - {\rm i} L)} \; v^2
\right]
\; .
\label{W0_7}
 \end{equation}
Using Eq.~(\ref{W0_3}) with $x_1=x_2=0$ together with Eq.~(\ref{W0_7}), the coherence between two points along the axis turns out to be described by the following function:
\begin{equation}
\begin{array}{rl}
W_z(\zeta_1,\zeta_2)= 
&
\beta(\zeta_1,\zeta_2)
\; e^{{\rm i} k L (\zeta_2-\zeta_1)}
\\
\\
& 
\times \displaystyle \int_{-\infty}^\infty p(v) \,
e^{- \pi^2 w_0^2\gamma(\zeta_1,\zeta_2) \, v^2}
{\rm d} v
\; ,
\end{array}
\label{W0_8}
\end{equation}
with
\begin{equation}
\beta(\zeta_1,\zeta_2)
 =
 \displaystyle\frac{\left| A \right|^2 }{\sqrt{(\zeta_1+{\rm i})(\zeta_2-{\rm i})}}
\; ,
\label{W0_9.1}
\end{equation}
and
\begin{equation}
\gamma(\zeta_1,\zeta_2)
=
\displaystyle\frac{2 \, \zeta_1 \zeta_2 + {\rm i} (\zeta_2-\zeta_1)}{(\zeta_1+{\rm i})(\zeta_2-{\rm i})}
\; ,
\label{W0_9.2}
\end{equation}
where the normalized coordinate $\zeta= z/L$ has been used.

In particular, the intensity at the point $\zeta$ along the axis is evaluated from Eqs.~(\ref{W0_8}), (\ref{W0_9.1}), and (\ref{W0_9.2}) on letting $\zeta_1=\zeta_2=\zeta$, which yields
\begin{equation}
\begin{array}{c}
I_z(\zeta)
=
\beta(\zeta,\zeta)
\displaystyle \int_{-\infty}^\infty p(v) \,
e^{- \pi^2  w_0^2 \, \gamma(\zeta,\zeta) \,  v^2}
{\rm d} v
\; ,
\end{array}
\label{W0_10}
\end{equation}
%
and the degree of coherence along the axis is evaluated as
\begin{equation}
\mu_z(\zeta_1,\zeta_2)
 =
\displaystyle\frac{W_z(\zeta_1,\zeta_2)}{\sqrt{I_z(\zeta_1)I_z(\zeta_2)}}
\; .
\label{W0_11.1}
\end{equation}

Equations~(\ref{W0_8})-(\ref{W0_9.2}) provide a simple analytical tool to evaluate the z-coherence of the beam radiated by a general SM source, with the only constraint that the source intensity has a Gaussian shape. Furthermore, as we shall see, in several cases of common interest the obtained expressions are quite simple and compact. Note that the case of a coherent Gaussian beam is obtained on taking $p(v)$ as a Dirac's delta function, for which both the integrals in Eqs.~(\ref{W0_8}) and (\ref{W0_10}) give one, and the CSD, that is
\begin{equation}
W_z(\zeta_1,\zeta_2)= 
\displaystyle\frac{\left| A \right|^2 \; e^{- {\rm i} k L (\zeta_1-\zeta_2)} }{\sqrt{(\zeta_1+{\rm i})(\zeta_2-{\rm i})}}
\; ,
\label{W0_11.3}
\end{equation}
factorizes, as it is expected for a coherent beam.

As a first example we take
\begin{equation}
\mu(x)
=
{\rm sinc} (x/\delta),
\label{ex2.1}
\end{equation}
where ${\rm sinc}(t)={\rm sin}(\pi t)/(\pi t)$ and $\delta$ is a parameter related to its width. The corresponding weight function, associated to $\mu$ by a Fourier transform, is
\begin{equation}
p(v)
=
\delta \, {\rm rect} (\delta v)
\; ,
\label{ex2.2}
\end{equation}
${\rm rect} (t)$ being the unit box, equal to 1 for $|t|\le 1/2$ and 0 otherwise. 
The non-negativity of $p(v)$ guarantees the correctness of the resulting CSD. 

The integral appearing in Eq.~(\ref{W0_8}), namely,
\begin{equation}
P(\gamma; q)
=
\displaystyle \int_{-\infty}^\infty p(v) \,
e^{- \pi^2 w_0^2 \,  \gamma \; v^2}
\, {\rm d} v
\; ,
\label{ex1.2.1}
\end{equation}
with this choice of $p$ is easily evaluated as
\begin{equation}
P(\gamma; q)
=
\displaystyle\frac{1}{\sqrt{q \gamma \pi}} \;
{\rm erf}\left[ \displaystyle\frac{\pi \sqrt{q \gamma}}{2}\right]
\; ,
\label{ex2.3}
\end{equation}
erf$[\cdot]$ being the error function~\cite{Gradshteyn65}.
The parameter $q$, defined as
\begin{equation}
q =
\left(\displaystyle\frac{w_0}{\delta}\right)^2
\; ,
\label{W0_12}
\end{equation}
has been introduced to give account of the global coherence features of the source and is the only parameter responsible for the z-coherence properties of the radiated beam.
Limiting cases of a spatially perfect coherent and a completely incoherent source are obtained for $q \to 0$ and $q \to \infty$, respectively, in which cases $P(\gamma; q) \to 1$ and $P(\gamma; q) \to 0$.

\begin{figure}[!ht]
	\centering
	\includegraphics[width=7 cm] {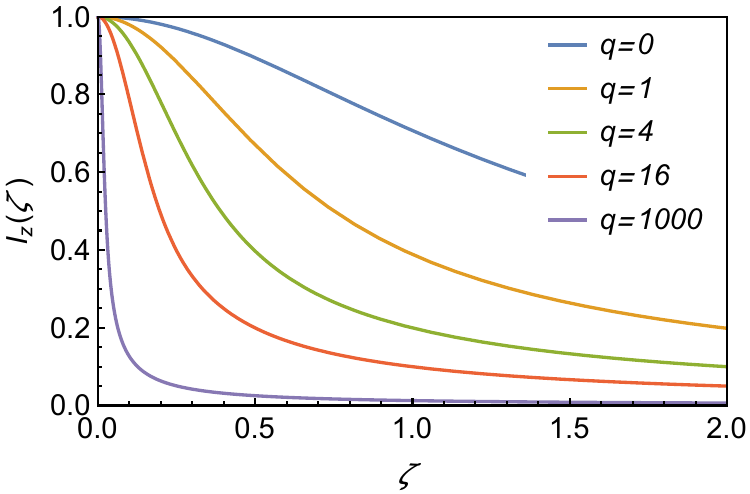}
	\caption{Axial intensity vs $\zeta$ for $\mu(x)$ given in Eq.~(\ref{ex2.1}) and increasing values of $q$: 0 (coherent source), 1, 4, 16, 1000 (almost incoherent source).}
	\label{ex1.inte}
\end{figure}

\begin{figure}[!ht]
\centering
\includegraphics[width=8 cm] {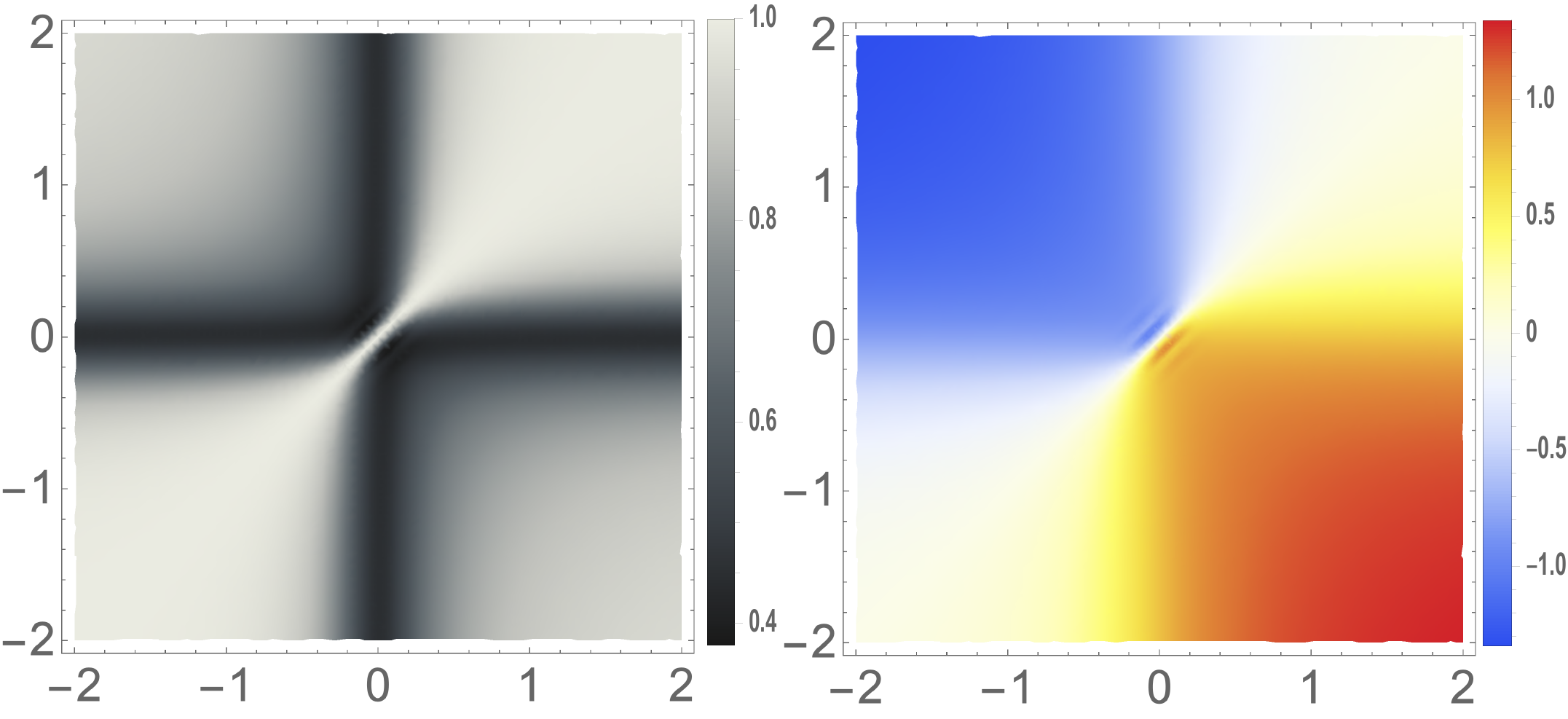}
\caption{Modulus  and argument  of $\mu_z$ across the plane ($\zeta_1$, $\zeta_2$)  for  for $\mu(x)$ given in Eq.~(\ref{ex2.1}) and $q=16$.}
\label{ex1.mu}
\end{figure}

\begin{figure}[!ht]
\begin{tabular}{c c}
\centering
 \includegraphics[width=4. cm]{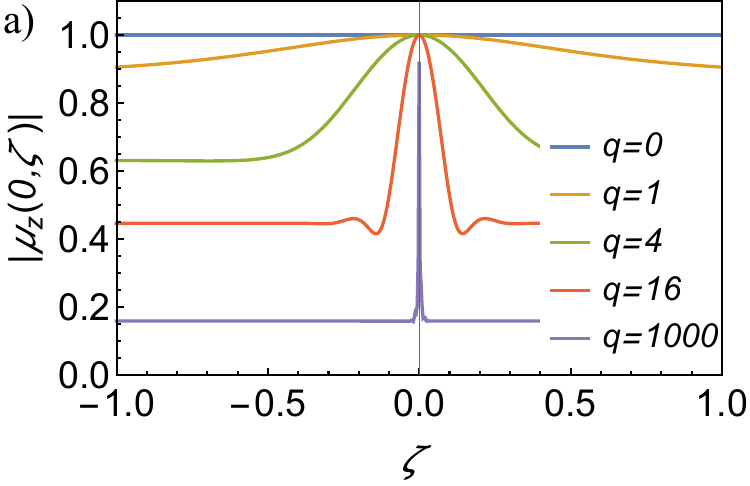} 
 & 
 \includegraphics[width=4. cm]{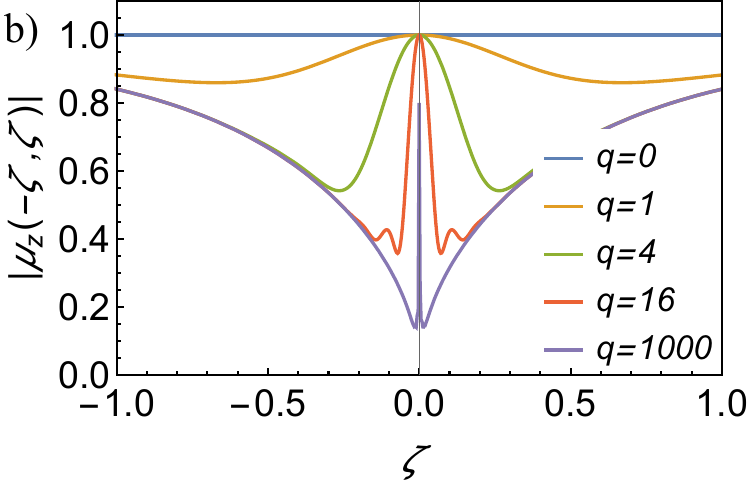} 
 \\
 \includegraphics[width=4. cm]{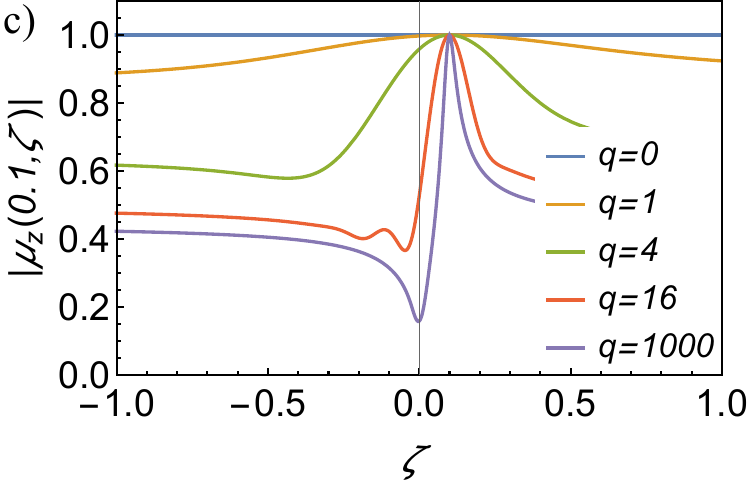} 
 &
 \includegraphics[width=4. cm]{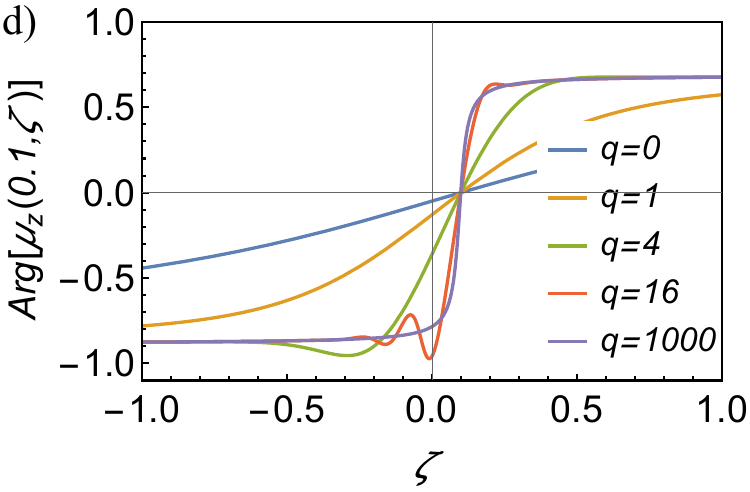} 
 \end{tabular}
\caption{Plots of $\mu_z(\zeta_1, \zeta_2)$ for $\mu(x)$ given in Eq.~(\ref{ex2.1}) for some choices of one of the two points as a function of the other: a) modulus of $\mu_z$ for $\zeta_1=0$ and $\zeta_2=\zeta$; b) modulus of $\mu_z$ for $\zeta_1=-\zeta$ and $\zeta_2=\zeta$; c) modulus of $\mu_z$ for $\zeta_1=0.1$ and $\zeta_2=\zeta$; d) argument of $\mu_z$ for $\zeta_1=0.1$ and $\zeta_2=\zeta$. 
}
\label{ex1.varimu}
\end{figure}

Figures from \ref{ex1.inte} to \ref{ex1.varimu} show some of the quantities calculated from the above expressions. We shall consider also cases in which one of the two points has a negative $\zeta$. Indeed, the Fresnel diffraction integral can be also used in backpropagation.

Figure~\ref{ex1.inte} shows the on-axis intensity as a function of the normalized axial coordinate for several values of $q$, while modulus and argument of $\mu_z(\zeta_1, \zeta_2)$ are shown in Fig.~\ref{ex1.mu} as 2D plots, for $q=16$. 

For a clearer visualization of the coherence between two axial points and the effect of the parameter $q$, Fig.~\ref{ex1.varimu} shows some plots of $\mu_z$ for different choices of one of the two points. In Fig.~\ref{ex1.varimu}a the absolute values is plotted as a function of $\zeta_1=0$ with $\zeta_2=\zeta$, while in Fig.~\ref{ex1.varimu}b we set $\zeta_1=-\zeta$ and $\zeta_2=\zeta$. 
From the first of these figures it is apparent that the correlation between the point on the source and a point very far away from it does not generally tend to zero but to some positive value. Such value can be evaluated from Eqs.(\ref{W0_8})--(\ref{W0_11.1}) and it turns out that, in this limit,
\begin{equation}
\left| \mu_z(0,\zeta) \right|
 \to
 \displaystyle\frac{P(1,q)}{\sqrt{P(2,q)}}
 \;.
 \label{W0_13}
\end{equation}
Such a result actually holds whatever the chosen $\mu(x)$  is.
With analogous considerations it can be shown that the curves in Fig.~\ref{ex1.varimu}b tend to 1 when $\zeta \to \pm \infty$. This can be understood on considering that, when the distance from the source is very large, both for positive and negative $z$, only one mode (the one propagating along the axis, i.e., the one with $v=0$) contributes significantly to the on-axis field.

Figures~\ref{ex1.varimu}c and \ref{ex1.varimu}d report modulus and argument of $\mu_z$ for $\zeta_1=0.1$ and $\zeta_2=\zeta$. Values of the parameter $q$ have been chosen as in Fig.~\ref{ex1.inte}. In the curves of in Fig.~\ref{ex1.varimu}d only the phase anomaly is plotted, i.e., the contribution $k L (\zeta_2-\zeta_1)$ present in Eq.~({\ref{W0_8}}) has been removed.


For the second example we take a DOC across the source of the form of a bilateral exponential, that is,
\begin{equation}
\mu(x)
=
e^{-|x/\delta|}
\; ,
\label{ex1.1}
\end{equation}
which corresponds to the weight function
\begin{equation}
p(v)
=
\displaystyle\frac{2 \delta}{1+(2 \pi \delta  v)^2}
\; ,
\label{ex1.2}
\end{equation}
which is positive for any $v$, thus ensuring that the resulting CSD is \emph{bona fide}.
The integral in Eq.~(\ref{ex1.2.1}) with this choice of $p$ is evaluated as
\begin{equation}
P(\gamma; q)
=
{\rm erfc}
\left[\sqrt{q \gamma}/2\right]
\,  e^{q \gamma/4}
\; ,
\label{ex1.3}
\end{equation}
where erfc$[\cdot]$ denotes the complementary error function~\cite{Gradshteyn65}. 
The same limiting values as for the previous case are obtained for high and small values of $q$, corresponding to an incoherent and to a perfectly coherent source. Figures analogous to the ones of the previous example could be plotted. 

Other possible choices of DOC (such as those with a Lorentzian shape, or like a sinc$^2$, or like a unit triangle, and so on) give rise to equally simple expressions of the propagated on-axis coherence function and  intensity, but we want to conclude with a whole class of functions which encompasses the Gaussian function. In such a way our results can be directly compared to the ones already obtained for a Gaussian Schell-model (GSM) source~\cite{Friberg1983}. Functions of this class are the so-called \emph{elegant} Hermite-Gaussian functions~\cite{Siegman1973} of even order, and we write the corresponding DOC as
\begin{equation}
\mu_m(x)
=
\displaystyle\frac{(-1)^m  m!}{(2m)!} \; {\rm H}_{2m}(x/\delta) \; e^{-(x/\delta)^2}
\; .
\label{ex3.1}
\end{equation}
Here, ${\rm H}_n(t)$ is the $n$th Hermite polynomial~\cite{Gradshteyn65} and the in front coefficient ensures that $\mu_m(0)=1 \; \forall m$. The zero-order member of this class is the standard Gaussian function.

That such functions can represent possible DOCs is certified by the corresponding weight functions, that are
\begin{equation}
p(v)
=
\displaystyle\frac{2^m}{(2m-1)!!} \, \delta \, \sqrt{\pi}  \, (\pi \delta v)^{2m} e^{-(\pi \delta v)^2}
\; ,
\label{ex3.2}
\end{equation}
which are never negative. Inserted into the integral in Eq.~(\ref{ex1.2.1}), they give
\begin{equation}
P(\gamma; q)
=
\left(1+q \gamma \right)^{-(m+1/2)}
\; .
\label{ex3.3}
\end{equation}
The latter provides, together with Eqs.~(\ref{W0_9.1}) and (\ref{W0_9.2}),
\begin{equation}
\begin{array}{c}
W_z(\zeta_1,\zeta_2)
= 
\displaystyle\frac{\left| A \right|^2 \; e^{{\rm i} k L (\zeta_2-\zeta_1)}}
{\sqrt{(\zeta_1+{\rm i})(\zeta_2-{\rm i})+ q (2 \, \zeta_1 \zeta_2 + {\rm i} (\zeta_2-\zeta_1))}}
\\
\\
\times \; \left[
\displaystyle\frac{(\zeta_1+{\rm i})(\zeta_2-{\rm i})}
{(\zeta_1+{\rm i})(\zeta_2-{\rm i})+ q (2 \, \zeta_1 \zeta_2 + {\rm i} (\zeta_2-\zeta_1))}
\right]^m
\; ,
\end{array}
\label{ex3.4}
\end{equation}
and, on letting $\zeta_1=\zeta_2=\zeta$,
\begin{equation}
I_z(\zeta)
=
\displaystyle\frac{\left| A \right|^2}{\sqrt{1+(1+2q)\zeta^2}}
\left[
\displaystyle\frac{1+\zeta^2}{1+(1+2q)\zeta^2}
\right]^m
\; .
\label{ex3.5}
\end{equation}

\begin{figure}[!ht]
	\centering
	\includegraphics[width=7 cm] {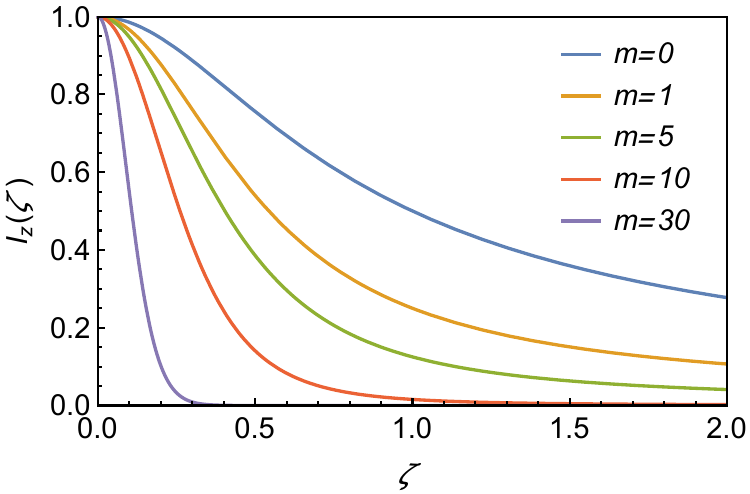}
	\caption{Axial intensity vs $\zeta$ given in Eq.~(\ref{ex3.5}) for $q=1$ and several values of $m$: 0 (GSM source), 1, 5, 10, 30.}
	\label{ex3.inte}
\end{figure}

\begin{figure}[!ht]
\centering
\includegraphics[width=8 cm] {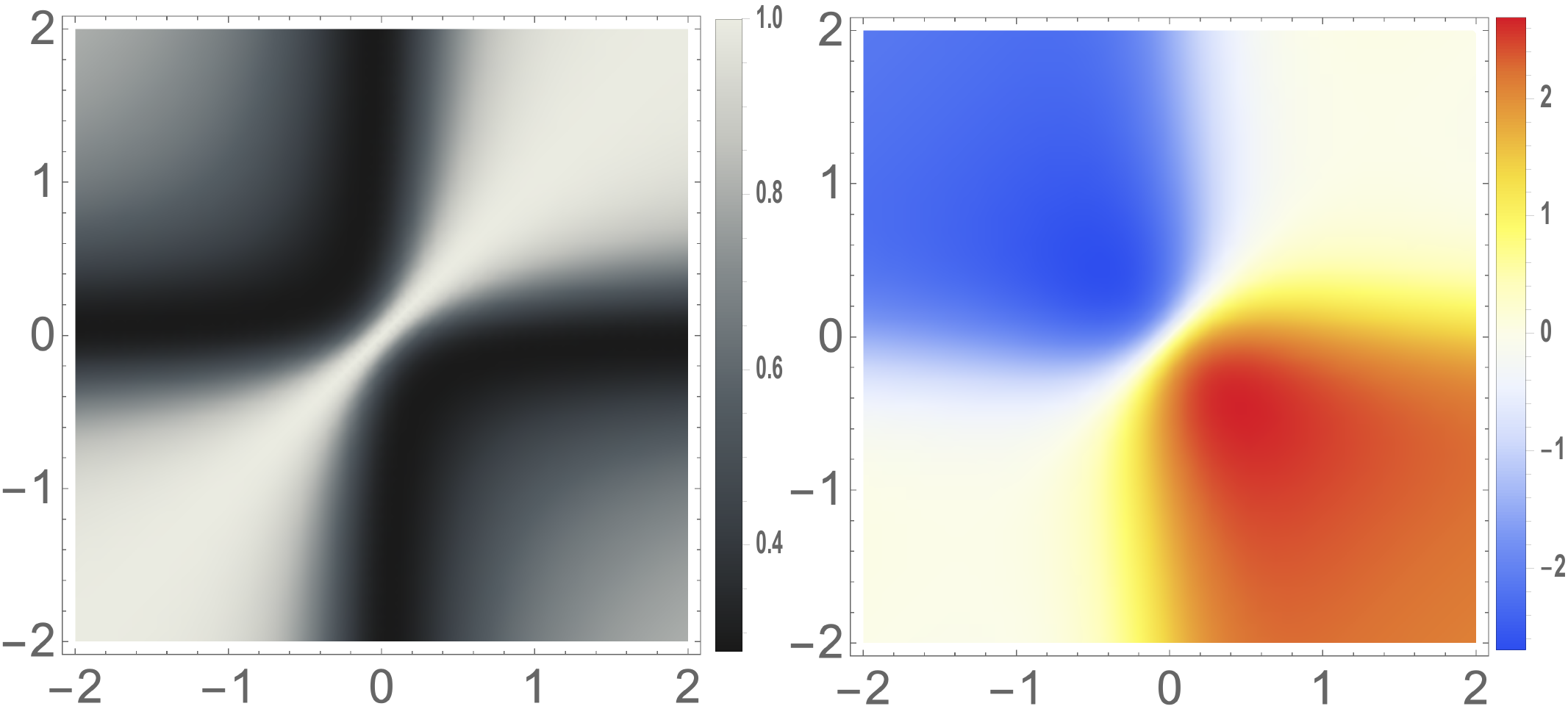}
\caption{Modulus  and argument  of $\mu_z$ across the plane ($\zeta_1$, $\zeta_2$) obtained from the CSD  in Eq.~(\ref{ex3.4}) for $m=2$ and $q=4$.}
\label{ex3.mu}
\end{figure}

\begin{figure}[!ht]
\begin{tabular}{c c}
\centering
 \includegraphics[width=4.1 cm]{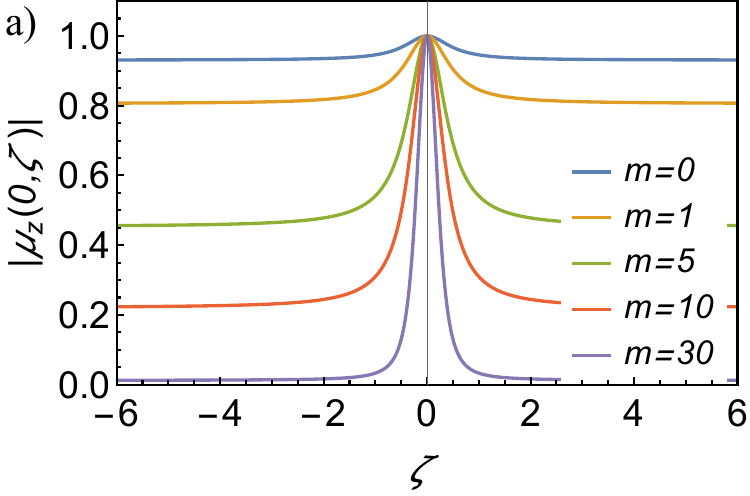} 
 &
 \includegraphics[width=4.1 cm]{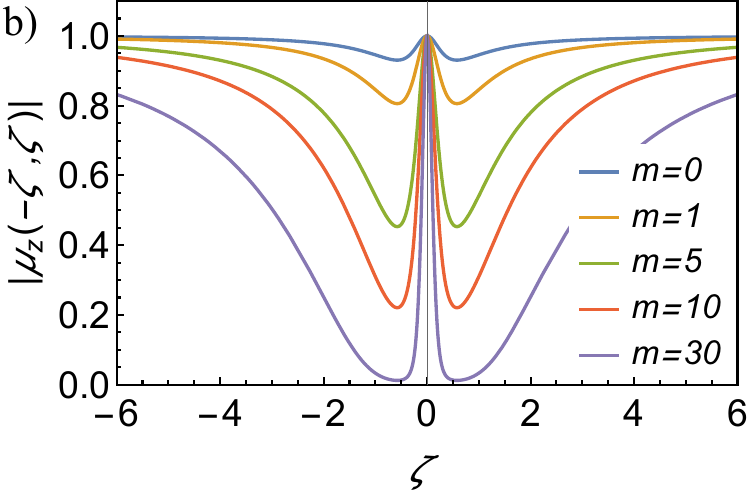} 
 \\
 \includegraphics[width=4.1 cm]{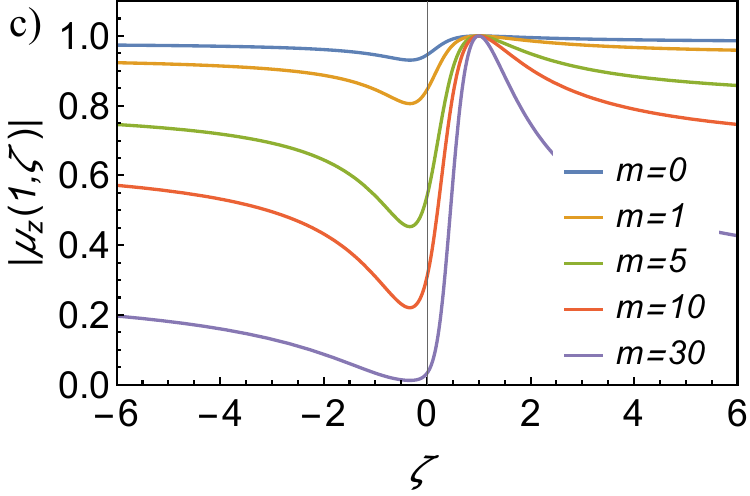} 
 &
 \includegraphics[width=4.1 cm]{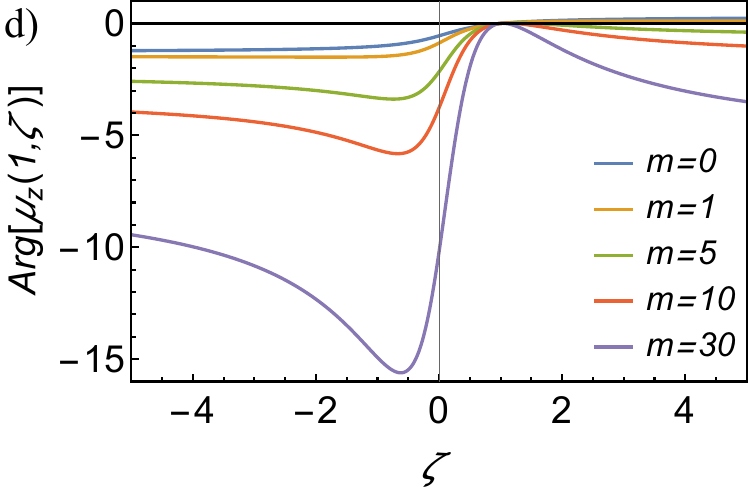} 
 \end{tabular}
\caption{Plots of $\mu_z(\zeta_1, \zeta_2)$ obtained from the CSD in Eq.~(\ref{ex3.4}) for some choices of one of the two points as a function of the other and $q=1$: a) modulus of $\mu_z$ for $\zeta_1=0$ and $\zeta_2=\zeta$; b) modulus of $\mu_z$ for $\zeta_1=-\zeta$ and $\zeta_2=\zeta$; c) modulus of $\mu_z$ for $\zeta_1=1$ and $\zeta_2=\zeta$; d) argument of $\mu_z$ for $\zeta_1=1$ and $\zeta_2=\zeta$. 
}
\label{ex3.varimu}
\end{figure}

The same quantities as for the previous example are shown in Figs.~\ref{ex3.inte}-\ref{ex3.varimu}, but now we kept fixed the value of $q$ (=1) and in each plot we consider different values of $m$, to put into evidence the role of the order of the HG mode.
It is seen that increasing the order corresponds, roughly speaking, to decreasing the coherence of the source. The reported behaviors, although with different functional forms, qualitatively agree with those obtained for the first example.

\vspace{.2cm}
In conclusion, in this Letter we presented a simple expression for the on-axis CSD of the beam radiated by a Schell-model source having general degree of coherence and Gaussian intensity profile. This result may have a significant impact in practical applications where the control of the z-coherence and/or the intensity along the axis is needed. In particular, the ability to fully control the z-coherence may help with overcoming the Rayleigh resolution limit~\cite{Liang2021,KorotkovaTSO2022} in imaging systems resolving the structures in the longitudinal direction, such as the OCT.

\vspace{.2cm}
{\bf Fundings.}  	Ministerio de Econom\'ia y Competitividad (PID2019-104268GB-C21)

\vspace{.2cm}
{\bf Disclosures.}  The authors declare no conflicts of interest.

\vspace{.2cm}
{\bf Data Availability.}  Data underlying the results presented in this paper are not publicly available at this time but may be obtained from the authors upon reasonable request.

\end{document}